**Solving the Schrodinger equation directly for a particle in one-dimensional periodic potentials**

Manoj K. Harbola

Department of physics

Indian Institute of Technology, Kanpur

Kanpur, India – 208016

**Abstract**

Solutions of time-independent Schrodinger equation for potentials periodic in space satisfy Bloch's theorem. The theorem has been used to obtain solutions of the Schrodinger equation for periodic systems by expanding them in terms of plane waves of appropriate wave-vectors and then diagonalising the resulting Hamiltonian matrix. However, this method can give exact solutions only if an infinite number of plane-waves are used in the wavefunction. Thus the Schrodinger equation has not been solved exactly even for one-dimensional infinitely large periodic systems despite the simplification that the theorem provides. On the other hand, solutions of one-dimensional Schrodinger equation for finite systems are obtained routinely by using standard numerical techniques to integrate the Schrodinger equation directly. In this paper, we extend this approach to infinitely large systems and use Bloch's theorem in a novel way to find direct exact numerical solutions of the Schrodinger equation for arbitrary one-dimensional periodic potentials. In addition to providing a different way of solving the Schrodinger equation for such systems, the simplicity of the algorithm renders it a great pedagogical value.

# I. Introduction

For a particle moving in an arbitrary potential, a *tour de force* method of obtaining its wavefunction and the associated eigenvalue is to diagonalise the corresponding Hamiltonian matrix in a suitable basis. Although the method provides highly accurate solutions of the Schrodinger equation, it can never be exact in principle. Thus it is always challenging - and charming - to actually solve the Schrodinger equation even for a restricted class of systems; the solution could be analytical or numerical. For example, exact analytical solutions exist for a particle moving in an infinitely large box or parabolic potential in one dimension and in the potential $v(r) = -\frac{Z}{r}$ or $v(r) = kr^2$, where $r$ is the distance from the origin, in three dimensions. Furthermore, for arbitrary potentials in one dimension and spherically symmetric potentials having arbitrary $r$ dependence in three dimensions, analytical solutions do not exist in general but bound-state solutions can be found by integrating the Schrodinger equation numerically; effectively three-dimensional spherically symmetric potentials also present themselves as one-dimensional problem since the angular solutions are known analytically. We sum up that exact bound-state solutions for a particle moving in a one-dimensional potential can always be found by solving the Schrodinger equation either analytically or numerically. However, even in one dimension there is a class of problems, viz., a particle moving in a periodic potential of infinite extent, for which direct solution of the Schrodinger equation has not been possible in general. This is in spite of the simplicity provided by the translational symmetry of the problem that renders the resulting wavefunctions a rather elegant and easy form given by Bloch's theorem.[1]

In this paper, we provide an algorithm that solves the Schrodinger equation for such systems exactly. This is the first time since Bloch's theorem was proved[1] in 1929 that it has been used to obtain a direct solution of the problem; the existing methods[2,3] of solution are built around expanding the wavefunction in terms of plane waves and diagonalising the resulting Hamiltonian matrix to obtain the solution. However, this method can never be exact (since that would require an infinite number of plane waves), although by including a large number of plane waves, the solution can be made highly accurate. Furthermore, it is discomforting to think that we have to diagonalise a large matrix to solve such a simple problem. It is against this background that the work presented in this paper assumes significance because it solves the Schrodinger equation for such kind of problems in a straightforward manner. Since the Schrodinger equation for a periodic potential is a special form of the Hill's equation[4], the work presented here provides a general method of solving the latter.

We conclude this section by discussing in the following paragraphs the Schrödinger equation for a particle moving in a one-dimensional periodic potential, Bloch's theorem for its solutions and the standard way of obtaining them. In the next section, we put Bloch's theorem in a form suitable for obtaining direct solutions of the Schrodinger equation. We then present the algorithm developed to solve the Schrodinger equation for these systems. We highlight the ease of its implementation and show its general character by applying it in section III to obtain (i) the energy bands for a particle moving in arbitrary one-dimensional periodic potentials and (ii) the photonic band structure for a one-dimensional photonic crystal. The paper is concluded by a discussion about the robustness of the algorithm and its pedagogic value.

**Solution of the Schrodinger equation for periodic potentials:** For a particle of mass $m$ moving in a periodic potential $v(x)$ of period $a$, such that $v(x + a) = v(x)$, the one dimensional Schrodinger equation

$$\left[-\frac{\hbar^2}{2m}\frac{d^2}{dx^2} + v(x)\right]\psi(x) = E\psi(x) \qquad (1)$$

has solutions $\psi(x)$ that satisfy the Bloch condition

$$\psi(x + a) = e^{ika}\psi(x) \qquad (2)$$

due to the translational symmetry of the problem. These solutions correctly lead to the probability density invariant under a finite translation of the system by *a*. Consequently, there is a good quantum number - the wavevector *k* - associated with a wavefunction and the corresponding eigenenergy is a function of this quantum number. Thus the wavefunction and eigenenergy are usually written as $\psi_k(x)$ and $E(k)$. By using Eq. (2), the wavefunction can be expanded in terms of plane waves as

$$\psi_k(x) = \sum_G C_{k+G} e^{i(k+G)x} \quad , \qquad (3)$$

where $G = n\left(\frac{2\pi}{a}\right); n = 0, \pm 1, \pm 2 \cdots$ are the reciprocal-space vectors; $C_{k+G}$ are unknown coefficients to be determined by the Schrodinger equation as described below.

The potential $v(x)$ can also be expanded in terms of the reciprocal space vectors as

$$v(x) = \sum_G v_G e^{iGx} \quad , \qquad (4a)$$

where the Fourier coefficients

$$v_G = \frac{1}{a}\int_0^a v(x) e^{-iGx} dx \quad . \qquad (4b)$$

Substitution of Eqs. (3) and (4) in the Schrodinger equation leads to the following set of equations [2, 3]

$$[\epsilon_{k+G} - E(k)]C_{k+G} + \sum_{G'} v_{G'} C_{k+G-G'} = 0 \qquad (5)$$

for the coefficients in Eq. (3) with $\epsilon_{k+G} = \frac{\hbar^2(k+G)^2}{2m}$. With a sufficiently large number of coefficients the equations above lead to accurate wavefunctions $\psi_k(x)$ and energies $E(k)$, approaching their exact values asymptotically. As is well known, the energy eigenvalues $E(k)$ form bands with the number of bands being equal to the number of plane waves used to solve the equations. However, because of the infinite number of coefficients involved, it is not possible to solve these equations in an exact manner. Furthermore, setting up of Eq. (5)

requires an additional step of calculating the Fourier coefficients of the periodic potential, which adds to the computational effort needed to obtain the solutions. On the other hand, if we are able to integrate the Schrodinger equation to obtain solutions satisfying the Bloch's condition, we will have the eigenfunctions and the eigenvalues for the system described by Eq. (1). How it is to be done is the subject of this paper.

## II. Solving the Schrodinger equation by integration

In general the wavefunction $\psi_k(x)$ is complex and can be written in terms of its real and imaginary components $\psi_k^R(x)$ and $\psi_k^I(x)$, respectively, as

$$\psi_k(x) = \psi_k^R(x) + i\psi_k^I(x) \quad . \tag{6}$$

Now the Bloch condition given by Eq. (2) can be written as

$$\psi_k^R(x) = \cos(ka)\,\psi_k^R(x+a) + \sin(ka)\,\psi_k^I(x+a) \tag{7a}$$

and

$$\psi_k^I(x) = -\sin(ka)\,\psi_k^R(x+a) + \cos(ka)\,\psi_k^I(x+a) \quad . \tag{7b}$$

We now employ Bloch's conditions, Eqs. (7a) and (7b), to obtain exact solutions of the Schrodinger equation by integrating it.

To develop the solutions, we will take the two boundaries of the unit cell to be located at $x = 0$ and at $x = a$. For a given energy, the Schrodinger equation for the real or imaginary part of the wavefunction is the same as that for the total wavefunction. Thus in principle both parts could be identical except that such a solution will not be consistent with the conditions expressed by Eqs. (7a) and (7b). To obtain the imaginary part $\psi_k^I(x)$ consistent with a given real part $\psi_k^R(x)$, we propose to use an iterative procedure as follows. For a given $k$ and $E$, we start by integrating the Schrodinger equation from $x = 0$ to $x = a$ with some arbitrarily chosen values of $\psi_k^R(0)$, $\frac{d\psi_k^R}{dx}(0)$, $\psi_k^I(0)$ and $\frac{d\psi_k^I}{dx}(0)$ to get a solution $\psi_k^R(x)$ and $\psi_k^I(x)$ [iteration number $n = 0$] and check if Eq. (7b) is satisfied by these solutions; if it is not, we keep $\psi_k^R(x)$ fixed and start with the new initial values

$$\psi_k^I(0)[n] = -\sin(ka)\,\psi_k^R(a)[n=0] + \cos(ka)\psi_k^I(a)[n-1] \ , \qquad (8a)$$

$$\frac{d\psi_k^I}{dx}(0)[n] = -\sin(ka)\frac{d\psi_k^R}{dx}(a)[n=0] + \cos(ka)\frac{d\psi_k^I}{dx}(a)[n-1] \qquad (8b)$$

for $\psi_k^I(x)$ and solve the Schrodinger equation again for it with these values. In the equations above, $n$ in square brackets indicates the iteration number. Note that during the iteration process, the real part of the wavefunction is kept fixed, indicated by $n = 0$. This is repeated until condition given by Eq. (7b) is satisfied within a specified numerical tolerance. Thus at the end of this process, we have the imaginary part of the wavefunction consistent with the Bloch condition (7b) for a given real part. We note that in the actual algorithm, it is better to use a mixture of the old (iteration number $n$-1) and new (iteration number $n$) initial values as the new initial value of the imaginary part for numerical stability of the iterative process.

To establish that the algorithm given above indeed gives the correct solutions, we take the example of a free particle which is described by the wavefunction

$$e^{i(kx+\varphi)} = \cos(kx + \varphi) + i\sin(kx + \varphi) \ . \qquad (9)$$

We start with $\psi_k^R(x) = \psi_k^I(x) = \cos(kx + \varphi)$ and apply the algorithm above to it with the energy $E(k) = \frac{\hbar^2 k^2}{2m}$. We have taken $a = 3.0$ so that $k$ ranges from $-\frac{\pi}{3}$ to $\frac{\pi}{3}$ and we do our calculations for $k = 1.0$ taking $\hbar = m = 1$. It is seen that after a few iterations $\psi_k^I(x)$ converges to $\sin(kx + \varphi)$ irrespective of the value of $\varphi$. Displayed in figure 1 are the real and imaginary parts of the wavefunction for $\varphi = 0$ and the corresponding probability density. Figure (1a) shows the three functions before achieving convergence (200 iterations); hence the probability density varies with $x$. Figure (1b) shows the same functions after convergence (1500 iterations) with the result that the probability density now becomes a constant. From the two figures the evolution of $\psi_k^I(x)$ during the iterative process is evident.

Having come up with an algorithm to find the imaginary part $\psi_k^I(x)$ of the wavefunction for a given real part, we next ask how do we find the real part $\psi_k^R(x)$ and the corresponding eigenenergy for a given $k$. To this end, we proceed as follows. For a given $k$, we choose an energy $E$ and integrate the Schrodinger equation to obtain $\psi_k^R(x)$ and $\psi_k^I(x)$ with some arbitrary values of $\psi_k^R(0)$, $\frac{d\psi_k^R}{dx}(0)$, $\psi_k^I(0)$ and $\frac{d\psi_k^I}{dx}(0)$, as was done earlier. We then run the iterative algorithm described above to find the final $\psi_k^I(x)$ consistent with Eq. (7b). We now ask if $\psi_k^R(x)$ satisfies Eq. (7a) with the $\psi_k^I(x)$ thus determined. If the answer is in the affirmative, we have found the energy eigenvalue $E(k)$ and the corresponding $\psi_k^R(x)$ and $\psi_k^I(x)$. If Eq. (7a) is not satisfied, we change the value of $E$ and go over the process described above until we find the solution within the prescribed tolerance limit. Thus the steps involved to solve the Schrodinger equation for one dimensional periodic potential can be recapitulated as follows:

(1) For a given $k$, choose a value of $E$;

(2) Find $\psi_k^R(x)$ and the corresponding $\psi_k^I(x)$ using the iterative algorithm given above;

(3) Check if $\psi_k^R(x)$ satisfies condition given by Eq. (7a). If yes, the solution has been found;

(4) If condition of Eq. (7a) is not satisfied, change the value of $E$ and go to step (1).

We point out that by integrating the Schrodinger equation directly, we also obtain the wavefunction in a straightforward manner. This is something that gives the method additional elegance and makes it pedagogically very useful.

We finally note a wavefunction and the corresponding eigenvalue for a given wavevector $k$ is also a solution for any wavevector $k + G$. This is because the $k$-dependence enters the solution via Eqs. (7a) and (7b) and if these equations are satisfied for a $k$, they are also satisfied by any $k + G$.

We now demonstrate our method by applying it in the next section to obtain energy bands for (i) the Kronig-Penny model, and (ii) a particle moving in a one dimensional model atomic chain. To demonstrate the generality of our method, the above mentioned quantum mechanical examples are followed by applying the method to another problem of interest, namely, the frequency bands for a one dimensional photonic band-gap material[5,6] made of slabs of two materials of different dielectric constants.

### III. Illustrative examples

**The Kronig-Penney model:** We start the demonstration of the algorithm developed above with the classic Kronig-Penny model consisting of the delta-function periodic potential $v(x) = \sum_{n=-\infty}^{\infty} V_0 \delta(x - na)$ that is periodic with period $a$. For convenience we take $\hbar = m = 1$ and the unit of length to be the Bohr radius $a_0$ of hydrogen atom. For a delta-function potential, the wavefunction is continuous across the potential but its derivative undergoes a step change proportional to $V_0$. Thus at $x = a$, we have

$$\psi_k^R(a^+) = \psi_k^R(a^-) \tag{10a}$$

$$\psi_k^I(a^+) = \psi_k^I(a^-) \tag{10b}$$

$$\frac{d\psi_k^R}{dx}(a^+) = \frac{d\psi_k^R}{dx}(a^-) + 2V_0 \psi_k^R(a) \tag{10c}$$

$$\frac{d\psi_k^I}{dx}(a^+) = \frac{d\psi_k^I}{dx}(a^-) + 2V_0 \psi_k^I(a) \tag{10d}$$

where in the last two equations we have used $\psi_k^R(a)$ and $\psi_k^I(a)$ for their values at $a^+$ or $a^-$ since they are the same. It is the values of these quantities at $a^+$ that are used in the algorithm above to obtain the solutions of the Schrodinger equation. For present calculations, we have chosen $a = 3$ and $V_0 = \frac{\pi}{2}$, which gives $aV_0 = \frac{3\pi}{2}$. This is the same value as used in the text-book example [2] of the model.

Shown in figure 2 are the energy bands in the first Brillouin zone, extending from $-\frac{\pi}{3}$ to $\frac{\pi}{3}$ (in units of $\frac{1}{a_0}$), obtained by solving the Schrodinger equation using the method described in section II. The energy values match perfectly with those given in the literature [2]. In figure 3a, we show the real and imaginary parts of the normalized wavefunction for $k = -0.26$ and the corresponding energy $E(k) = 1.980$ in the second band. This wavefunction was obtained starting with the initial values of $\psi_k^R(0) = 1.0$ and $\frac{d\psi_k^R}{dx}(0) = 1.0$; we are not giving the values of $\psi_k^I(0)$ and $\frac{d\psi_k^I}{dx}(0)$ since they become irrelevant after the iterative process to find the imaginary part converges. The corresponding probability density is also shown. To test the robustness of the method, we repeated our calculation with a different set of initial conditions $\psi_k^R(0) = -10.0$ and $\frac{d\psi_k^R}{dx}(0) = 3.0$. It leads to exactly the same energy-band diagram as given in figure 2. We also present in figure 3b the wavefunction and the corresponding probability density for this calculation for the same *k* and *E(k)* as in figure 3a; we see that the probability density matches exactly with that shown in figure 3a. This lends credibility to the correctness of our method.

We note that in applying our method, we have two arbitrary constants, viz., the values of $\psi_k^R(0)$ and $\frac{d\psi_k^R}{dx}(0)$ in the solution. This is consistent with the text-book solution [2] of the problem, where also the solution contains two arbitrary constants. These two constant could be the amplitude and the phase of the undetermined coefficient. Finally, we note that the Schrodinger equation for the Kronig-Penney model has also been solved[7] in reciprocal space.

**Chain of one-dimensional atoms:** As the second example, we calculate the energy-bands for a chain of one-dimensional atoms. In each unit cell, the atom is situated in the middle and its potential energy (for the cell between $x = 0$ and $x = a$) is modeled as

$$v(x) = \frac{V_0}{\sqrt{\left(x - \frac{a}{2}\right)^2 + b^2}} \tag{11}$$

We have chosen $V_0 = 1.0, a = 3.0$ and $b = 0.2$. Again for simplicity we take $\hbar = m = 1$ and the unit of distance to be the Bohr radius of hydrogen atom. In figure 4 we show the corresponding band-diagram in the first Brillouin zone and in figure 5 the normalized wavefunction and the probability density for $k = 0.5$ and $E(k) = -0.218$ in the second band. These solutions have been obtained starting with $\psi_k^R(0) = 4.0$ and $\frac{d\psi_k^R}{dx}(0) = -1.0$. Notice that the lowest band is almost flat as expected. Finally, as for the Kronig-Penney model, we also tested our algorithm by running it with different initial values of $\psi_k^R(0)$ and $\frac{d\psi_k^R}{dx}(0)$. To our satisfaction, the resulting band-diagram and the probability density always comes out to be the same.

**Photonic bands:** To see if the method proposed herein is general in its scope, we next apply it to find the photonic band-structure of an infinitely large periodic one-dimensional material made of slabs of two materials with different dielectric constants. The equation for an electromagnetic wave propagating normal to the slabs (direction taken to be the x-direction) is

$$\left(\frac{d^2}{dx^2} + \frac{\omega^2(k)\epsilon(x)}{c^2}\right)\mathcal{E} = 0 \tag{12}$$

Here $\epsilon(x)$ is the dielectric constant varying periodically as a function of $x$ and $\mathcal{E}$ is the electric field. Because of the periodicity of $\epsilon(x)$, solutions of Eq. (12) also satisfy the Bloch's condition and therefore our method can be applied to obtain the allowed $\omega(k)$ as a function of $k$.

We have performed our calculations for the slab geometry with the slabs being taken to be of materials with dielectric constant 12.0 and 1.0 with their thickness being 2.0 and 1.0, respectively. Thus the period of the system is $a = 3.0$. Further, for simplicity we have taken $c = 1$. With unit of length being taken as $10^{-6}$m, the units of $k$ and $\omega$ are $10^6 \text{m}^{-1}$ and $3 \times 10^{14} \text{ rad.s}^{-1}$, respectively. The structure of the resulting photonic bands is shown in figure 6. As is evident from the diagram, the lowest photonic band is linear near $k = 0$ while the other bands have zero slope at that point. These results match with other calculations with similar geometry[4].

Although the results presented above have been obtained using a simple variation of the refractive index, the method developed in this paper is equally applicable to any general variation. In addition, our method also gives the corresponding electric field[6] in a straightforward manner.

## IV.    Discussion

In the preceding sections, we have presented a new general method of solving the Schrodinger equation with periodic external potentials, or similar second-order differential equations, using the Bloch's condition in a novel way. The physically observable quantities, such as the energy and the probability density, obtained are independent of the initial values used to start integration of the equation. Thus the method proposed is robust and consistent with the fact that Bloch's condition alone is sufficient to provide solutions of such problems. This can further be understood as follows. Solutions of the Bloch form can be written as $\psi_k(x) = Ae^{i(kx+\theta)}u_k(x)$, where $u_k(x)$ is a function periodic in space with period $a$ and $A$ and $\theta$ are two arbitrary real constants. The values of $A$ and $\theta$ determine the wavefunction and its slope at $x = 0$. Thus we have the freedom to choose these; this is precisely what has been done in our approach by taking arbitrary values of $\psi_k^R(0)$ and $\frac{d\psi_k^R}{dx}(0)$. As for the imaginary part of the wavefunction, it is fixed in accordance with the real part irrespective of the initial values of $\psi_k^I(0)$ and $\frac{d\psi_k^I}{dx}(0)$. Finally, because of the ease of its implementation, the method given in this paper can be easily included in a graduate or an advanced undergraduate course on quantum mechanics, computational physics or condensed matter physics.

Possible further studies using the method presented in this paper are: (i) study of tight binding wavefunctions for a solid, (ii) extending it to two and three dimensions as well as to a system of interacting particles, (iii) applying the method to solve the time-dependent Schrodinger equation (TDSE) with potentials periodic in time. As is well known, solutions of TDSE with time-periodic potentials are similar[8, 9] to Bloch's solutions except that the periodicity is now in time instead of space.

**Acknowledgement:** I thank Raj Chhabra for detailed comments on the manuscript.

# References

[1] F. Bloch, "Uber die Quantenmechanik der elektronen in kristallgittern," Z. Phys. **52** 555-600 (1929).

[2] C. Kittel, *Introduction to Solid-State Physics*, 7$^{th}$ edition (John-Wiley, Singapore, 1996) pp. 173-196.

[3] N. Ashcroft and N.D. Mermin, S*olid-State Physics*, 1$^{st}$ edition (Saunders College, Philadelphia, 1976) pp. 131-150.

[4] W. Magnus and S. Winkler, *Hill's equation* (Interscience publishers, a division of John Wiley and Sons, New York, 1966) pp. 3-46.

[5] I. Nusinsky and A.A. Hardy, "Band-gap analysis of one-dimensional photonic crystals and conditions for gap closing," Phys. Rev. B **73** (12), 125104-1-125104-9 (2006).

[6] G.V. Morozov and D.W.L. Sprung, "Floquet-Bloch waves in one-dimensional photonic crystals," Eur. Phys. Lett. **69** (5), 54005-p1-54005-p6 (2011).

[7] S. Singh, "Kronig-Penney model in reciprocal lattice space," Am. J. Phys. **51** (2), 179 (1983).

[8] P.W. Langhoff, S.T. Epstein and M. Karplus, "Aspects of time-dependent perturbation theory," Rev. Mod. Phys. **44** (3), 602-644 (1972).

[9] H. Sambe, "Steady states and quasienergies of a quantum-mechanical system in an oscillating field," Phys. Rev. A **7** (6), 2203-2213 (1973).

# Figure captions

**Figures 1a and 1b:** Figure 1a shows the real part $\psi_{real}(x) = \cos(x)$ of the free-particle wave function and its imaginary part $\psi_{imag}(x)$ obtained by applying iterative process discussed in the text before convergence has been achieved. The corresponding probability density is also shown. In figure 1b we show the same quantities after achieving convergence.

**Figure 2:** Band structure for the Kronig-Penney model with delta function potential. The band-diagram has been drawn by interpolating between the k-points (shown in the figure by dots) at which calculations were performed. For accurate interpolation near the zone-boundaries, calculations were also performed for points outside the first Brillouin zone and these points were used in the interpolation.

**Figures 3a and 3b:** In figure 3a we show the real part $\psi_{real}(x)$ of the normalized wave function for the Kronig-Penney model and its imaginary part $\psi_{imag}(x)$ for $k = -0.26$ and the corresponding energy $E(k) = 1.980$ in the second band, obtained by integrating the Schrodinger equation with the initial values of $\psi_k^R(0) = 1.0$ and $\frac{d\psi_k^R}{dx}(0) = 1.0$. The corresponding probability density is also shown. Figure 3b shows the same quantities for a different set of initial values $\psi_k^R(0) = -10.0$ and $\frac{d\psi_k^R}{dx}(0) = 3.0$ for the real part of the wavefunction.

**Figure 4:** Band structure for an atomic-chain potential with the atomic potential given by Eq. (11). The band- diagram has been drawn by interpolating between the k-points (shown in the figure by dots) at which calculations were performed. For accurate interpolation near the zone-boundaries, calculations were also performed for points outside the first Brillouin zone and these points were used in the interpolation.

**Figures 5:** Real part $\psi_{real}(x)$ of the normalized wave function for the atomic-chain potential and its imaginary part $\psi_{imag}(x)$ for $k = 0.5$ and the corresponding energy $E(k) = -0.218$ in the second band, obtained by integrating the Schrodinger equation with the initial values of $\psi_k^R(0) = 4.0$ and $\frac{d\psi_k^R}{dx}(0) = -1.0$. The corresponding probability density is also shown.

**Figure 6:** Photonic band structure for a one-dimensional photonic crystal made of stacks of dielectric slabs of different dielectric constants.

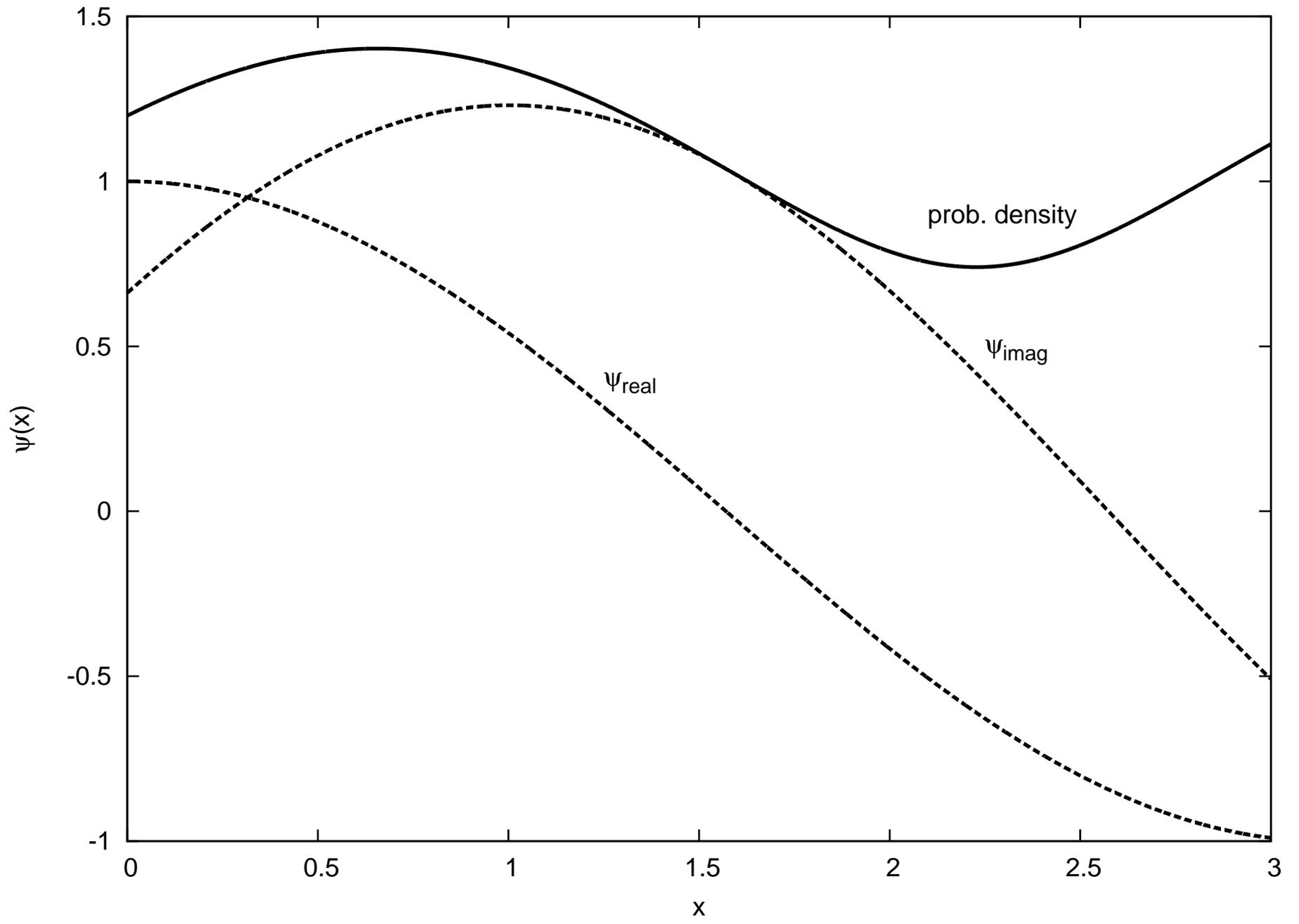

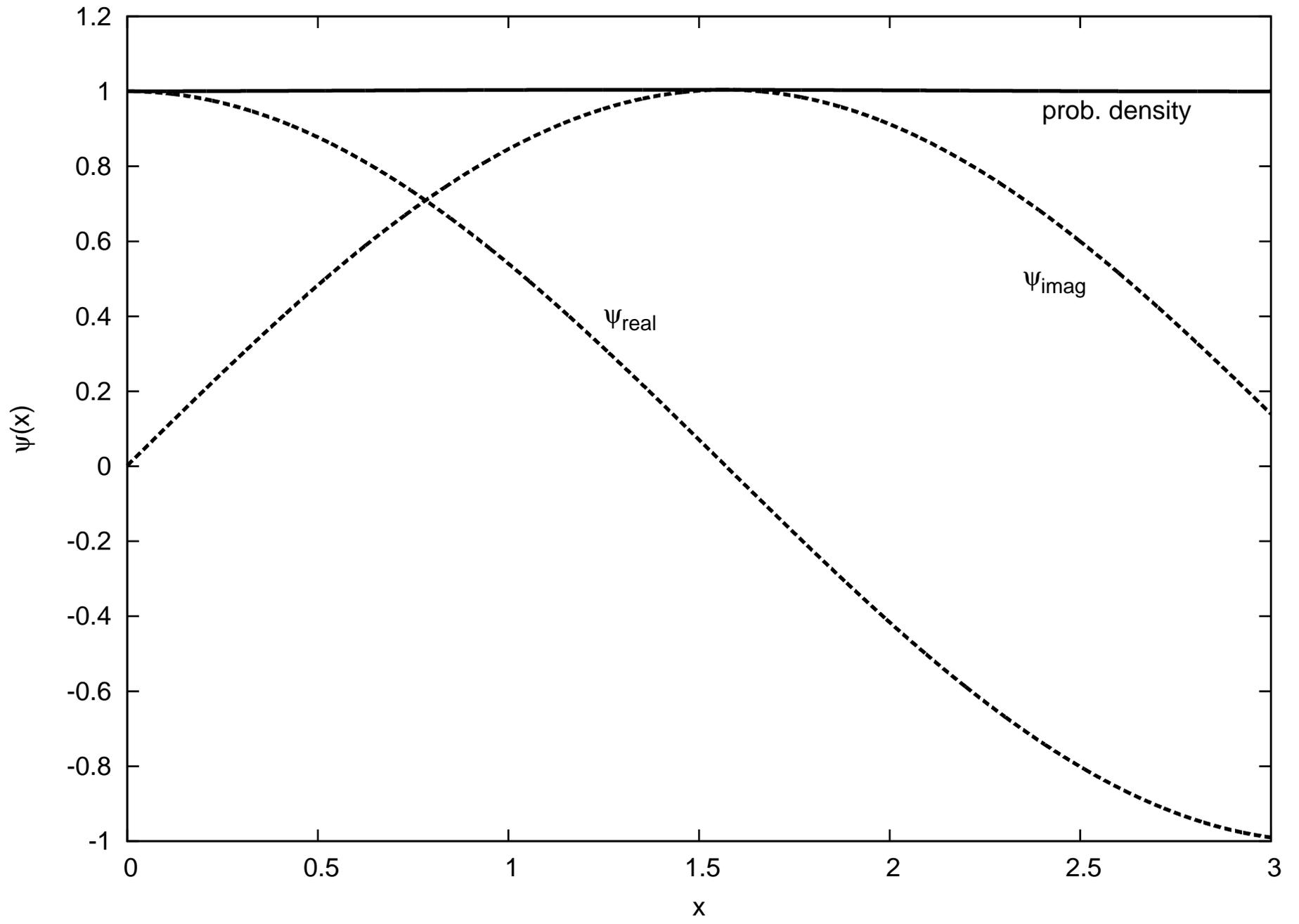

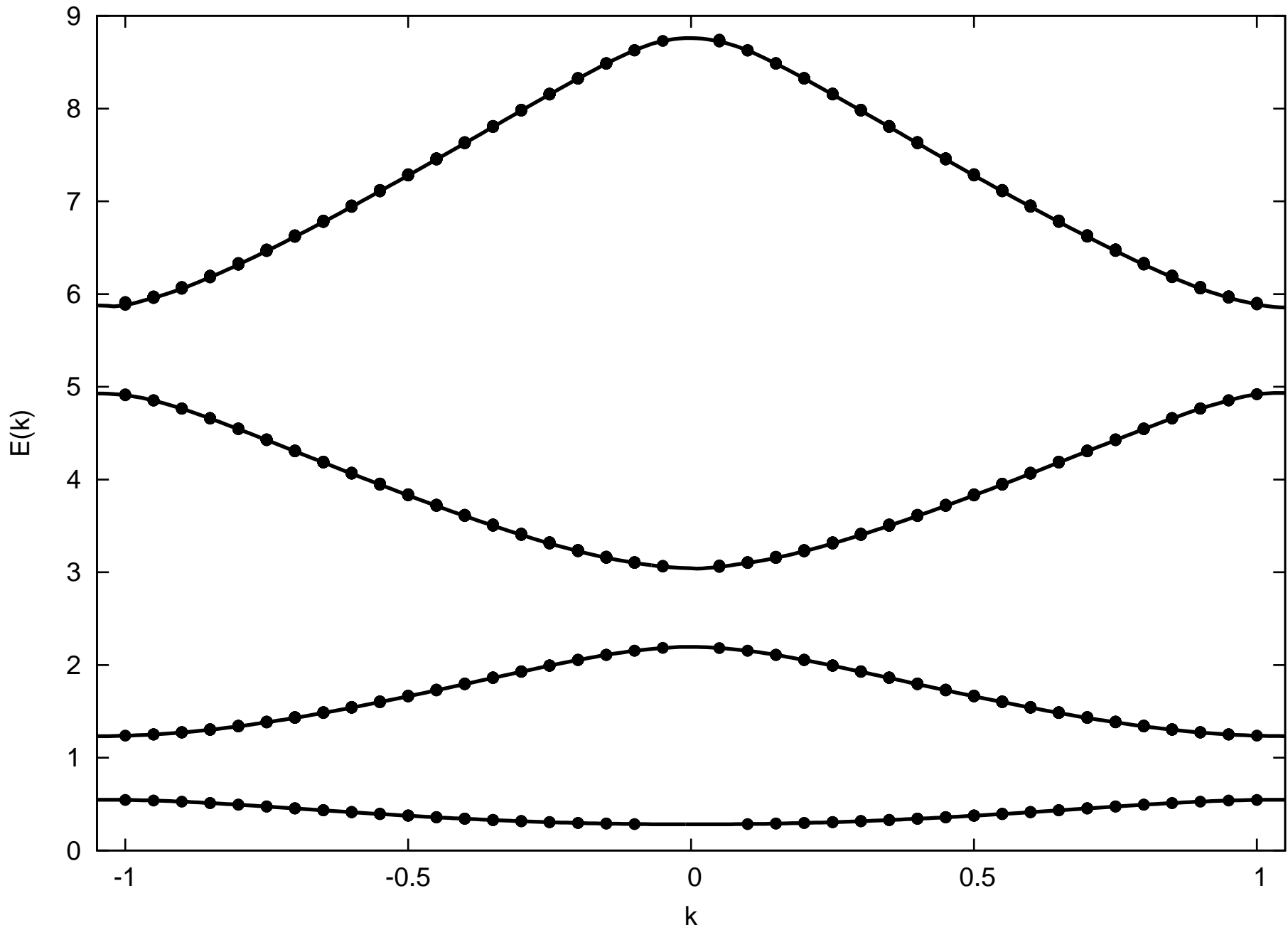

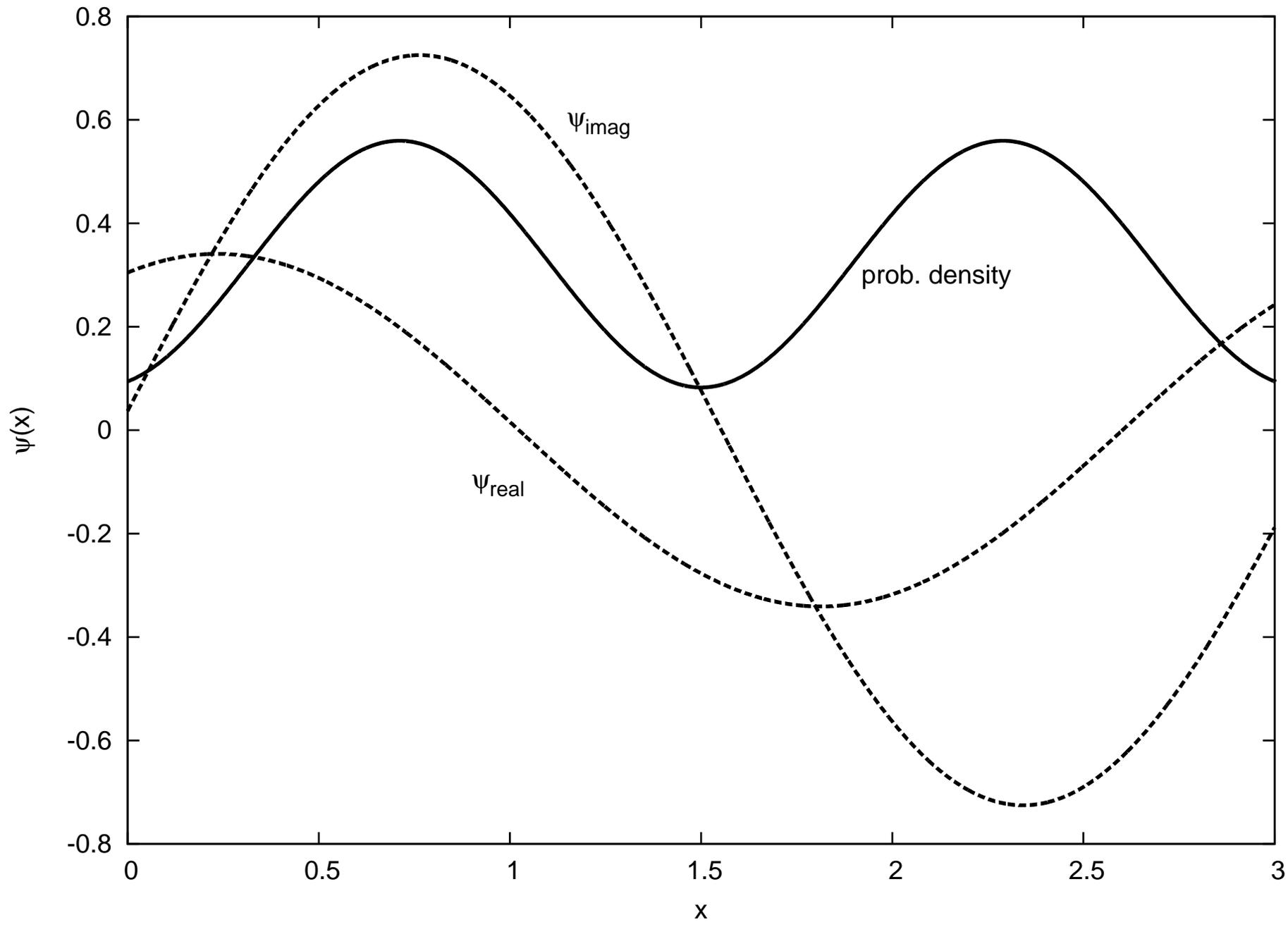

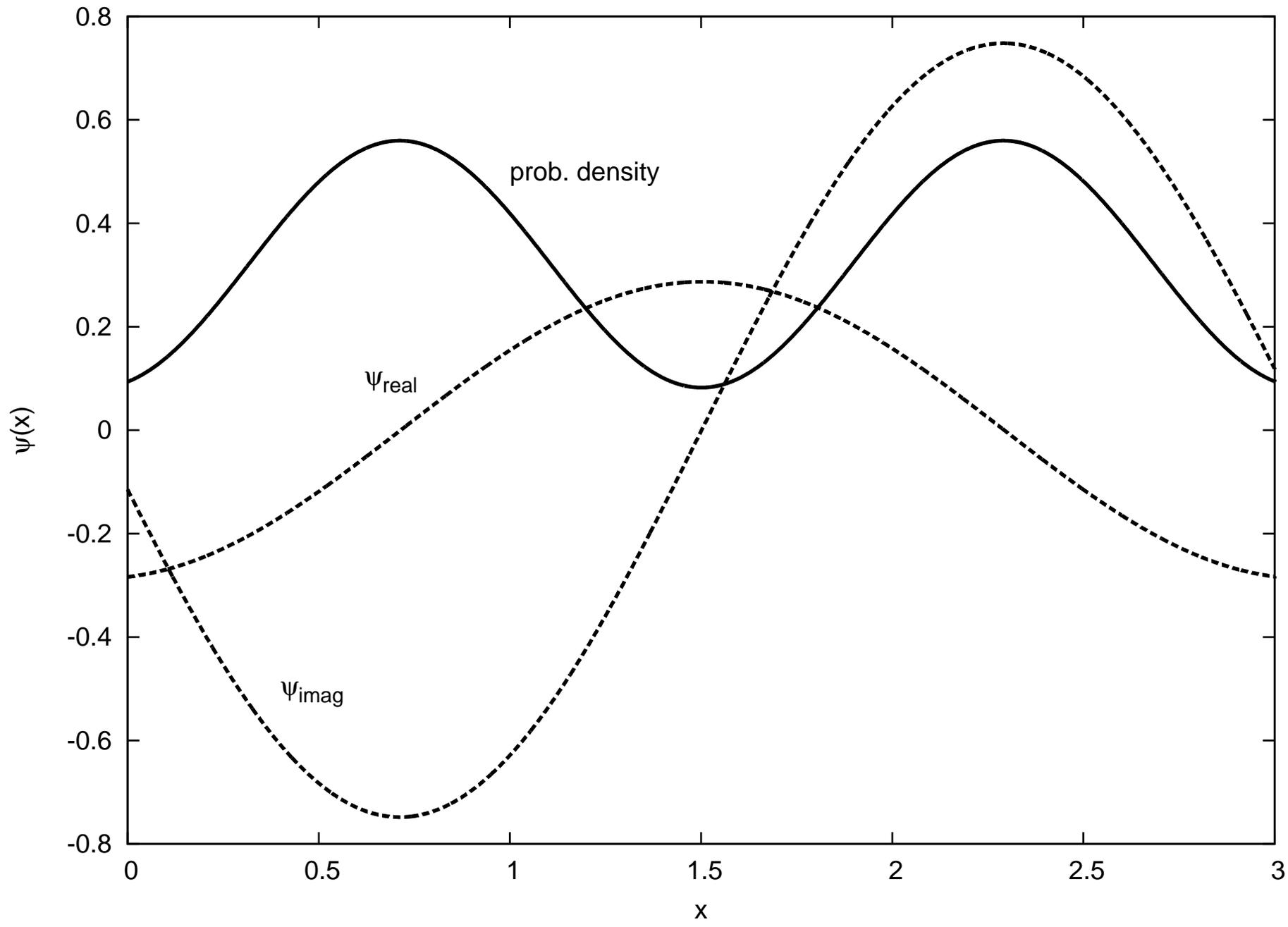

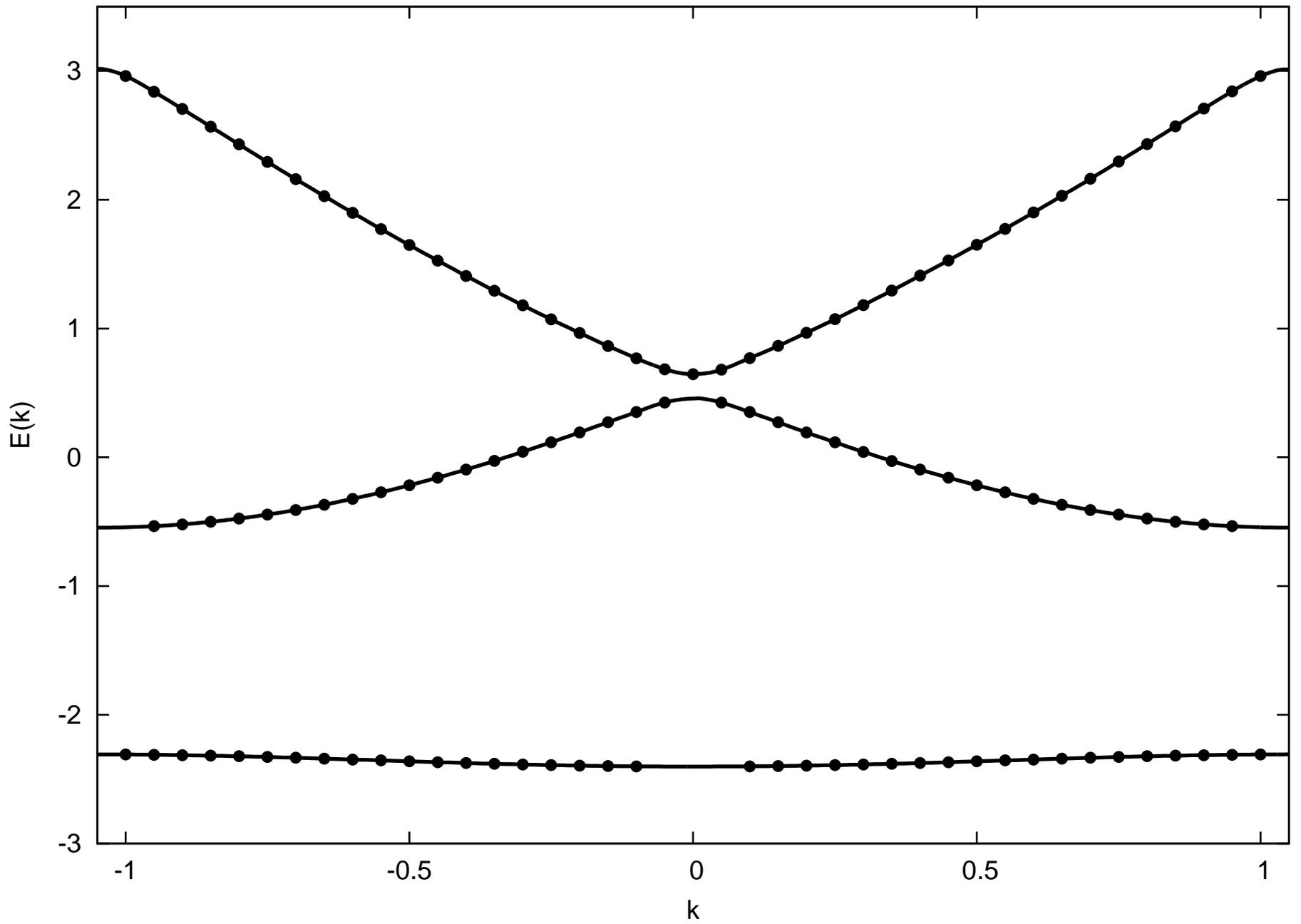

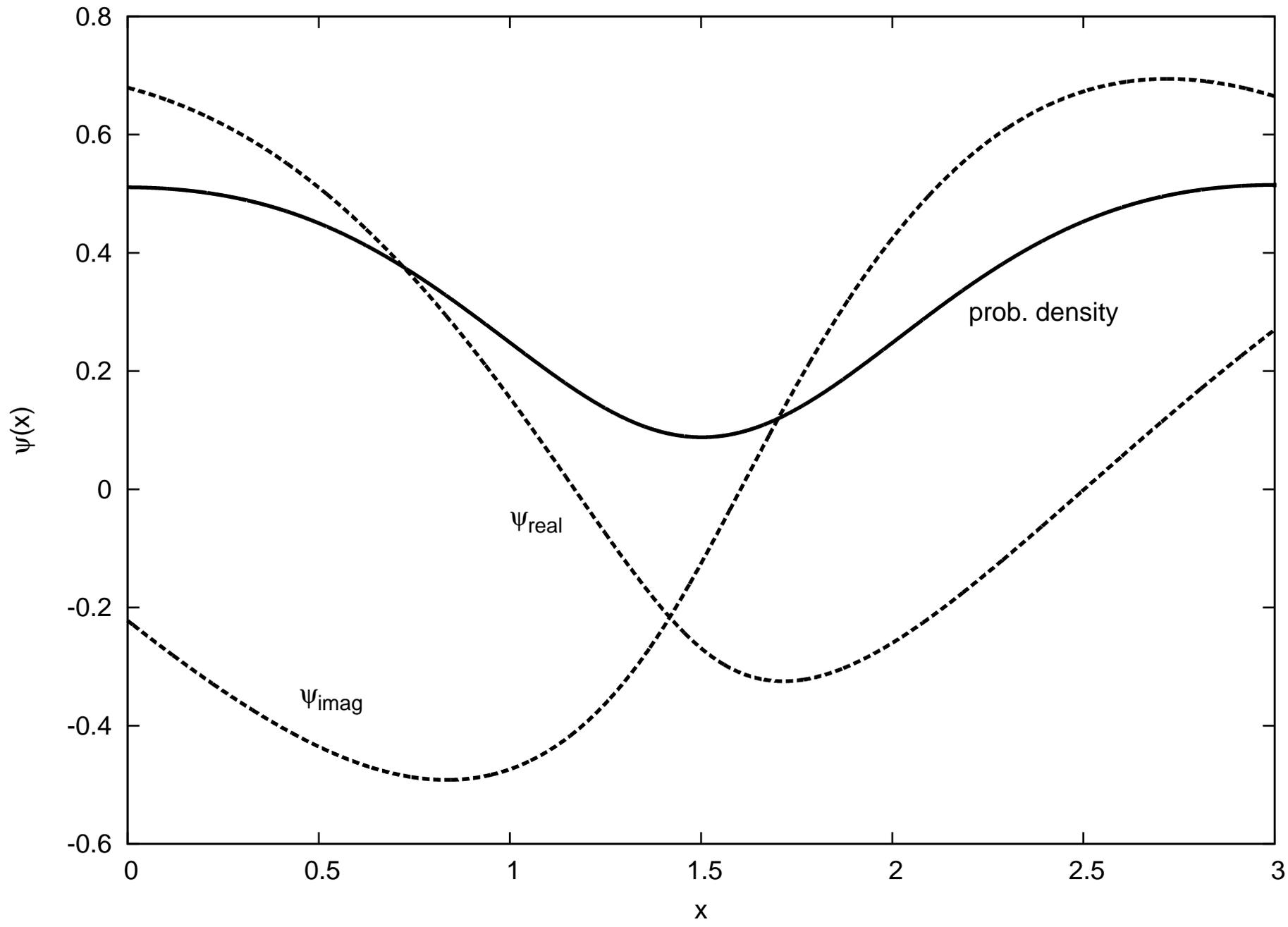

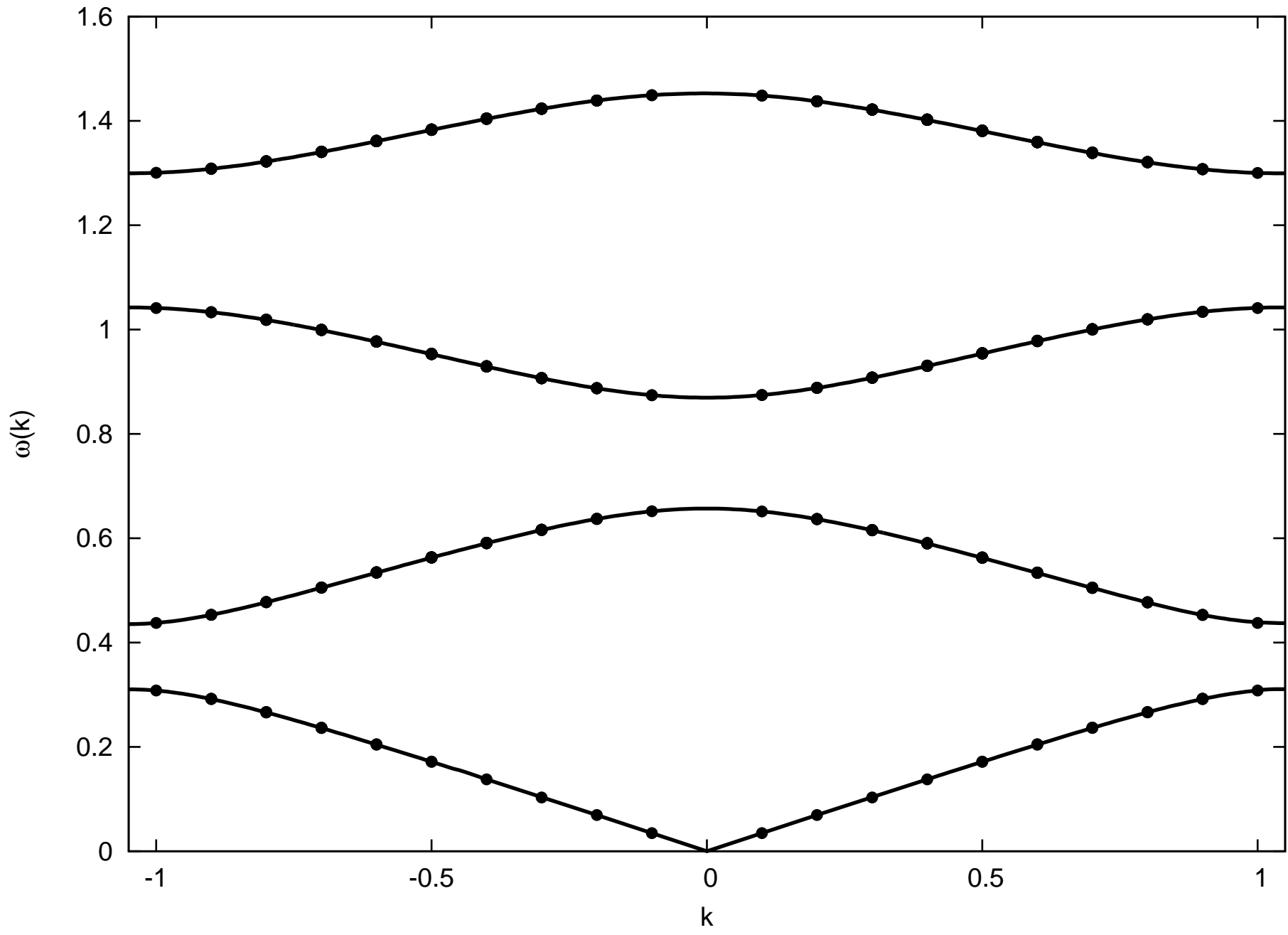